\documentclass[10pt]{article}
\newcommand{\intsum}{\hspace{1mm}\int\hspace*{-4mm}\sum}

\usepackage{amssymb}
\usepackage{latexsym} 
\usepackage{mathptm}
\usepackage{epsfig}
\usepackage[dvips]{color}
\definecolor{gruen}{rgb}{0.3,0.0,0.2}
\definecolor{gelb}{rgb}{0.0,0.3,0.2}

\begin{document}

\markboth{Sabine Hossenfelder}{Large Extra Dimensions and the Minimal Length}

\title{Large Extra Dimensions and the Minimal Length}
\author{Sabine Hossenfelder\thanks{sabine@physics.arizona.edu}\\
\small{Department of Physics}\\ 
\small{University of Arizona}\\
\small{1118 East 4th Street}\\ 
\small{Tucson, AZ 85721, USA}}

\date{}
\maketitle

\begin{abstract}
\textcolor{gruen}{Large extra dimensions lower the Planck scale to values soon accessible. 
Motivated by String Theory, the models of large extra dimensions predict a vast number of
new effects in the energy range of the lowered Planck scale, among them the production of TeV-mass 
black holes and gravitons. But not only is the Planck scale the energy scale at which effects of 
modified gravity become important. String Theory as well as non-commutative quantum mechanics 
suggest that the Planck length acts as a minimal length in nature, providing a natural ultraviolet 
cutoff and a limit to the possible resolution of spacetime. Within the extra dimensional scenario, 
the minimal length effects thus become important in the same energy range in which the effective extra 
dimensional models predict new physics. We examine a model which includes the minimal length into the 
extra dimensional extension of the Standard Model.}
 
\textcolor{gelb}{To appear in the Proceedings of the conference {\sl Physics@LHC}, 
Vienna, Austria, July 2004.}
\end{abstract}

\section{The Minimal Length Scale}
 
It was in the 5th century b.c. that Demokrit postulated a smallest particle out of which matter is build. 
He called it an ``atom''. In Greek, the prefix ``a'' means ``not'' and the word ``tomos'' means cut. 
Thus, atomos or atom means uncuttable or undividable. 
2500 years later, we know that not only the atom is dividable, but also is the atomic nucleus. 
The nucleus is itself a composite of neutrons and protons and further progress in science 
has revealed that even the neutrons and protons have a substructure. Is there an end to 
this or will the quarks and gluons turn out to be non-fundamental too?
  
An answer to this can not be given yet, but judging from one of the most promising 
candidates for an unified theory -- String Theory -- there is indeed a smallest 
possible resolution of spacetime\cite{Gross:1987ar}. This is not surprising since the success of String 
Theory is due to the very reason that the extension of strings is finite. The scale for 
this expected minimal length is given by the string scale which is close to the Planck-scale 
{$l_{\rm p}$}.

The occurrence of this minimal length scale has to be expected from very general reasons, not only
in String Theory but in all theories at high energies which attempt to include effects of
quantum gravity. A minimal length can be found in quantum loop gravity and non-commutative geometries
as well. This can be understood by the following phenomenological argument. 

Usually, every sample under investigation can be resolved by using beams of an energy 
high enough to assure the Compton-wavelength {$\lambda$} is below the size of the probe. 
The smaller the sample, the higher the energy must become and thus, the bigger the collider.
When the size of the probe should be as small as the Planck-length, the energy needed for 
the beam would be about Planck-mass. The Planck-mass, $m_{\rm p}\approx 10^{16}$~TeV, is the mass 
at which contributions of quantum gravity 
are expected to become important and at which curvature of space-time becomes non-negligible. 
The perturbation of space-time 
causes an uncertainty in addition to the usual uncertainty in quantum mechanics. This additional uncertainty 
increases with energy and {makes it impossible to probe distances below the Planck-length}.

The Planck-length which is derived from the Standard Model (SM) of physics is {$\approx 10^{-20}$ fm} and 
thus far out of reach for future experiments. But this depressing fact looks completely different if we work 
within the model of Large eXtra Dimensions ({\sc LXD}s). Here, the energy scale of quantum gravity can be considerably
lowered -- which means, the minimal length will be raised.

Effects of a minimal length scale have been examined before in various approaches and the analytical
properties of the resulting theories have been investigated closely\cite{Kempf:1994su,minilengthgeneral,hydrogen}. 
In the scenario without extra dimensions, the derived modifications are important mainly for structure formation
and the early universe\cite{earlyuniverse}.

\section{Large Extra Dimensions}

During the last decade several models using compactified {\sc LXD}s as an
additional assumption to the quantum field theories of the SM have
been proposed. These effective models are motivated by String Theory and provide
us with an useful description to predict first effects beyond the SM.
They do not claim to be a theory of first principles or a candidate for a grand
unification. Instead, the {\sc LXD}s allow us to compute testable results which can in turn
help us to gain insights about the underlying theory.
 
There are different ways to build a model of extra dimensional space-time. Here, we want to
mention only the most common ones:
\begin{enumerate}
\item The {\sc ADD}-model proposed by Arkani-Hamed, Dimopoulos and Dvali\cite{add} adds $d$ extra
spacelike dimensions without curvature, in general each of them compactified to the same radius $R$. All 
SM particles are confined to our brane, while gravitons are allowed to propagate freely in the bulk. \label{ADD}
\item The setting of the model from Randall and Sundrum\cite{rs1,rs2} is a 5-dimensional spacetime with
an non-factorizable geometry. The solution for the metric is found by analyzing the solution of Einsteins 
field equations with an energy density on our brane, where the SM particles live. In the type I model\cite{rs1} 
the extra dimension is compactified, in the type II model\cite{rs2} it is infinite.
\item Within the model of universal extra dimensions\cite{uxds}
all particles (or in some extensions, only bosons) can propagate in the 
whole multi-dimen\-sional spacetime. The extra dimensions are compactified on an orbifold to 
reproduce SM gauge degrees of freedom.
\end{enumerate}

In the following we will focus on the model (\ref{ADD}).
The radius $R$ of these extra dimensions typically
lies in the range mm to $10^3$~fm for $d$ from $2$ to $7$, or the inverse radius 
$1/R$ lies in energy range eV to MeV, respectively.
Due to the compactification, momenta in the direction of the {\sc LXD}s can only occur in quantized steps 
{$\propto 1/R$}. This yields an infinite number of equally spaced excitations, the so called Kaluza-Klein-Tower.
The existence of {\sc LXD}s leads to
a theoretical prescription which lowers the scale of quantum gravity down to values comparable to the
scales  of the SM interactions. We will denote
this new fundamental scale by {$M_{\rm f}$}. 
In order to solve the hierarchy problem, 
{$M_{\rm f}$} should be in the range of $\approx$~TeV. In the models with {\sc LXD}s, 
{first effects of quantum gravity then occur at energies} {$\approx M_{\rm f}$}. They
would be observable in soon future at the {\sc LHC}. Some of the predicted effects 
are\cite{Kanti:2004nr}: Production of gravitons, 
modifications due to virtual graviton exchange, production of black holes.

For self-consistence, the models further have to consider the fact that a lowering of the fundamental
scale leads to a raising of the minimal length. Within the model of {\sc LXD}s not only effects of quantum 
gravity occur at lowered energies but so do the effects of the minimal length scale.

\section{A Model for the Minimal Length}
 
To incorporate the notion of a minimal length into ordinary quantum field theory we will apply a
simple model which has been worked out in detail in\cite{Hossenfelder:2003jz,Hossenfelder:2004up}. 

\begin{figure}
\hspace*{-1cm}\epsfig{figure=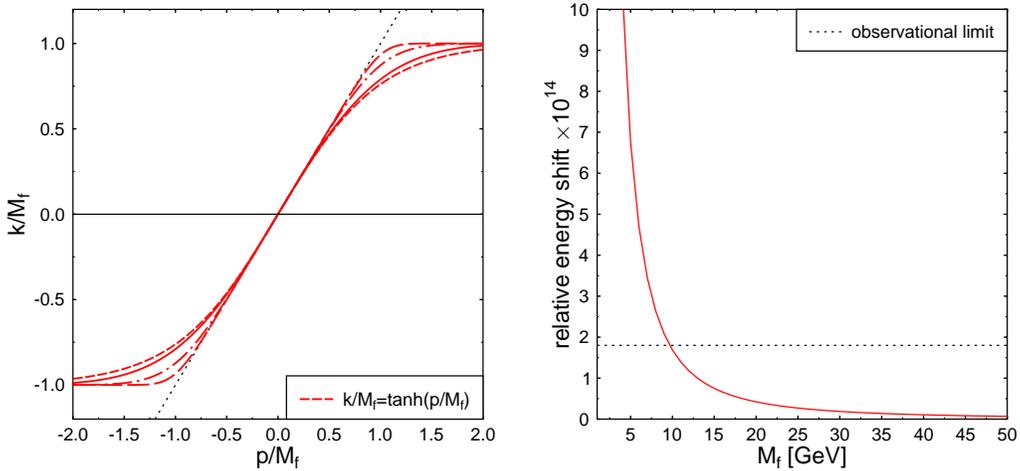, width=14cm}\vspace*{-0cm}
\caption{The left plot shows various possible choices for the functional relation $k(p)$ used in the
literature. The thin
dotted line corresponds to $k= M_{\rm f}\tanh(p/M_{\rm f})$.
The right plot \protect\cite{Hossenfelder:2004ze} shows the relative shift of the S1-S2 energy level in the
hydrogen atom as a function of the minimal length. The horizontal line shows the current observational
accuracy which is $\approx 1.8\times10^{-14}$ \protect\cite{Niering}.
\label{fig1}}
\end{figure}

We assume, no matter how 
much we increase the momentum $p$ of a particle, we can never
decrease its wavelength below some minimal length $L_{\mathrm f}$ or, equivalently, we can never increase
its wave-vector $k$ above $M_{\mathrm f}$. Thus, the relation between the 
momentum $p$ and the wave vector $k$ is no longer linear $p=k$ but a 
function
.
This function $k(p)$ has to fulfill the following properties:
\begin{enumerate}
\item For energies much smaller than the new scale we reproduce the linear relation: 
for $p \ll M_{\mathrm f}$ we have $p \approx k$ \label{limitsmallp}.
\item It is an odd function (because of parity) and $k$ is collinear to $p$.
\item The function asymptotically approaches the upper bound $M_{\mathrm f}$. \label{upperbound}
\end{enumerate} 
An example for a function fulfilling the above properties is $k= M_{\rm f}\tanh(p/M_{\rm f})$, which
is 
pictured in  Fig. \ref{fig1}, left.  
It is further assumes that $L_{\rm f} \ll R$, such
that the spacing of the Kaluza-Klein excitations compared to energy scales $M_{\rm f}$ 
becomes almost continuous and we can use an integration instead of summing up the KK-tower.

Lorentz-covariance is not added to the above list, as the proposed model can not provide conservation
of this symmetry. This is easy to see if we imagine an observer who is boosted relative to the minimal length.
He then would observe a contracted minimal length which would be even smaller than the minimal length. To
resolve this problem it might be inevitable to modify the Lorentz-transformation. Several attempts to construct
such transformations have been made\cite{Amelino-Camelia:2002wr}.
We will assume that
$p$ is a Lorentz vector and aim to express all quantities in terms of $p$.

The quantization of these relations is straightforward and follows the usual procedure. 
The commutators between the corresponding operators $\hat{k}$ and $\hat{x}$ 
remain in the standard form. 
Using the well known commutation relations and inserting the functional relation between the
wave vector and the momentum then yields the modified commutator for the momentum 
\begin{eqnarray}  
[\hat x_i,\hat k_j]={\mathrm i } \delta_{ij}\quad\Rightarrow\quad 
[\,\hat{x}_i,\hat{p}_j]&=& + {\rm i} \frac{\partial p_i}{\partial k_j} \quad.
\end{eqnarray}
This results in the generalized uncertainty principle ({\sc GUP)}
\begin{eqnarray} \label{gu}
\Delta p_i \Delta x_j \geq \frac{1}{2}  \Bigg| \left\langle \frac{\partial p_i}{\partial k_j} 
\right\rangle \Bigg| \quad, 
\end{eqnarray}
which reflects the fact that by construction it is not possible to resolve space-time distances
arbitrarily well. Since $p(k)$ gets asymptotically constant its derivative $\partial p/ \partial k$
drops to zero and the uncertainty in (\ref{gu}) increases for high energies. 
The behavior of our particles thus agrees with those of the strings found by Gross as mentioned above.

Since $k=k(p)$ we have for the eigenvectors $\hat{p}(\hat{k})\vert k \rangle= p(k)\vert k \rangle$ and 
so $\vert k \rangle \propto \vert p(k) \rangle$. We could now add that both sets 
of eigenvectors have to be a complete orthonormal system and 
therefore $\langle k' \vert k \rangle = \delta(k-k')$, 
$\langle p' \vert p \rangle = \delta(p-p')$. 
This seems to be a reasonable choice at 
first sight, since  $\vert k \rangle$ is known from the low energy regime. 
Unfortunately, now the normalization of the states is different 
because $k$ is restricted to the Brillouin zone
$-1/L_{\mathrm f}$ to $1/L_{\mathrm f}$. 
To avoid the need to recalculate normalization factors, we  
choose the $\vert p(k) \rangle$ to be identical to 
the $\vert k \rangle$. Following the proposal of \cite{Kempf:1994su} this yields then
a modification of the measure in momentum space.
In the presence of $d$ {\sc LXD}s with radii $R$, the eigenfunctions are then normalized to\cite{Hossenfelder:2004up}
\begin{eqnarray} \label{norm}
\langle p'(k') \vert p(k) \rangle &=& (2 \pi )^{3+d} \delta(k'_x-k_x) \delta_{k_y'k_y} R^d \nonumber \\
&=& (2 \pi )^{3+d}  \delta(p'_x-p_x) \Bigg| \frac{\partial p_i}{\partial k_j} \Bigg| \delta_{p_y'p_y}R^d
\quad,
\end{eqnarray}
where the functional determinant of the relation is responsible for an extra factor accompanying the
$\delta$-functions. The completeness relation of the modes takes the form
\begin{eqnarray} 
\intsum \frac{{\mathrm d}^3 k_x}{(2 \pi)^{d+3}} \frac{\langle k' \vert k \rangle}{N} = 
  R^d {\mathrm{Vol}}_d(p_y) 
\quad,
\end{eqnarray}
where ${\mathrm{Vol}}_d(p_y)$ denotes the Volume of the $d$-dimensional momentum space. 
To avoid a new normalization $N$ of 
the eigenfunctions we take the factors into the integral by a redefinition of the measure in momentum space 
\begin{eqnarray} \label{rescalevolume3}
{{\mathrm d}^{d+3} k} \rightarrow {{\mathrm d}^{d+3} p}  \Bigg| 
\frac{\partial k_i}{\partial p_j} 
\Bigg| \frac{1}{{\mathrm{Vol}}_d(p_y) R^d} \quad.
\end{eqnarray}
This redefinition has a physical interpretation because we expect the momentum 
space to be squeezed at high momentum values and weighted less. 
With use of an expansion of the tanh for high energies we have  
$\partial p / \partial k  \approx \exp\left( - |p|/M_{\rm f}\right)$ 
and so we can draw the important conclusion that the momentum measure is exponentially squeezed at high energies.

One can now retrace the usual steps and derive equations of motion in quantum mechanics and quantum field theory.
It can be shown\cite{Hossenfelder:2003jz,Hossenfelder:2004up} that the general replacement $p\to p(k)$ and the new measure Eq. (\ref{rescalevolume3}) turn out
to be a very simple recipe to rewrite the usual equations and Feynman rules.
So, the the quantization of the energy-momentum relation yields the modified Klein-Gordon Equation
and the Dirac Equation
\begin{eqnarray}  
\eta^{\mu\nu} 
\hat{p}_\nu(k) \hat{p}_\mu(k)  \psi =  m^2 \psi \quad,\quad (p\hspace{-1.6mm}/(k)-m)\psi&=&0  
\end{eqnarray}
where $p$ is now a function of $k$. 

\section{Observables}

These relations lead to modifications of the familiar equations in quantum mechanics and can be used to 
make predictions for effects that should arise from the existence of a minimal length scale. 
The momentum operator in position representation can be derived and with this, the modified 
Schr\"odinger equation:  
\begin{eqnarray}  
\hat{\vec{p}} = -{\rm i} \; \hbar  {\nabla} \left(1 - L_f^2 \Delta \right)
\quad,\quad
\hat{H} &=& 
- \frac{\hbar^2}{2m}    \left[ \nabla  \left(1 -  L_f^2  \Delta \right)  \right]^2  + V(r) \quad.
\end{eqnarray}

\subsection{The Hydrogen Atom}
For the hydrogen atom we insert  
$V(r)=e^2/r$ and find a shift of the energy levels from the old values {$E_n$} to the new values $\tilde{E}_n $. 
With the current accuracy of experiments for the Hydrogen S1-S2 transition, we find a very weak 
constraint on the new scale (see also Fig. \ref{fig1}, right)
\begin{eqnarray}  
\tilde{E}_n \approx E_n \left( 1 - \frac{4}{3} \frac{m_{\rm e}}{M_{\rm f}^2} \frac{E_0}{n^2} \right) \quad \Rightarrow \quad
M_{\rm f} \approx > 10 {\mbox{ GeV}} \quad.
\end{eqnarray}

These results with the new model are in agreement with those derived within
other models featuring the minimal length \cite{hydrogen}.

\subsection{The Muon $g-2$}

To derive useful predictions we have to look at high energy experiments 
or high precision observables, such as the magnetic moment of the muon {$g-2$} 
\cite{Harbach:2003qz}.
  
A particles energy in a magnetic field {$B$} depends on its spin. 
Energy levels that are degenerated for free particles split up in the presence of a field.
In a static, homogeneous magnetic field, the expectation value of the spin vector will
perform a precession around the direction of the field. The rotation frequency is proportional to 
the strength of the field and the magnetic moment of the particle and so can be used to measure 
the magnetic moment.

The value of $g$ is modified by self energy corrections in quantum field theory.
{Modifications from the minimal length should be important even at the classical level} and 
occur in the {\sc QED}-range. The experimental value for the magnetic moment of the muon is 
known today to extremely high precision.
The modifications arising from the minimal length can be derived by coupling the particle to the 
electromagnetic field 
 {$K_\nu \to k_\nu - e A_\nu$} in the usual way    
\begin{eqnarray}   
\omega \vert\psi\rangle \approx \gamma^0\left( \gamma^i\hat{K}_i +\frac{m}{\hbar}\right)\left(1- \frac{\hbar^2 
\hat{K}^i\hat{K}_i + m^2}{M_{\rm f}^2}\right)\vert\psi\rangle \quad.
\end{eqnarray}
Analyzing this equation, one finds a constraint on the new scale in the range of the constraints 
from {\sc LXD}s: 
{$M_{\rm f} \approx 0.67$ TeV}. 
 
\subsection{Tree Level Processes}

To go for the high energy observables, we examine how the minimal length influences cross-sections. We
compute the modifications of {$A+B \to X+Y$} cross-sections for {\sc QED} at tree level and find 
that an extra factor arises. 
 
\begin{figure}[t]
\hspace*{-1cm}\epsfig{figure=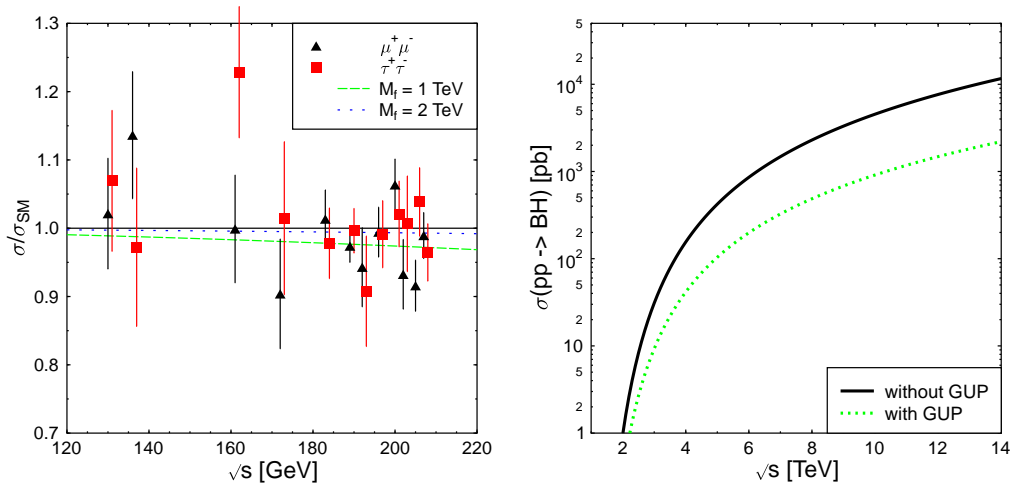, width=14cm}\vspace*{-0cm}
\caption{The left plot\protect\cite{Hossenfelder:2003jz} shows the energy dependence of the ratio from the new 
total cross-section value to
the SM cross-section for fermion pair 
production for different values of the minimal length. 
The data points are taken from \protect\cite{lep}.
The right plot\protect\cite{Hossenfelder:2004ze} shows the total cross-section 
for black hole production with minimal length $1/$TeV 
as a function of the center of mass energy $\sqrt{s}$. The ratio of the 
total cross section with and without {\sc GUP} for the expected {\sc LHC}-energy $\sqrt{s}=14$~TeV is 
$\approx 0.19$.
\label{fig2}}
\end{figure}

The dominant contribution is the squeezing-factor from the measure of momentum space which lowers the
phase space of the outgoing particles. This yields a cross-section
that is below the SM value. Applying this result to fermion pair production processes 
{$e^+e^- \to \mu^+ \mu^-$} or {$e^+e^- \to \tau^+ \tau^-$}, resp., we find that 
the modified cross-sections are in agreement with the data for a minimal length in the range 
{$L_f \approx 10^{-4}$ fm}, see Fig \ref{fig2}, left. 

\subsection{ And More}

The effect of the squeezed momentum space does also {modify 
predictions from the {\sc LXD}-scenario}. Since the
modifications get important at energies close to the new scale, the predicted graviton and black hole production is 
strongly influenced. The dilepton production under inclusion of virtual graviton exchange in hadron collisions 
has been examined in the minimal length scenario in Ref.\cite{Bhattacharyya:2004dy}.
 
Black hole production is less probable \cite{Hossenfelder:2004ze} since the approach of the partons to 
distances below the Schwarzschild-radius needs higher energies within the minimal length scenario.
Figure \ref{fig2}, right, shows the total cross-section of the black hole production. For {\sc LHC} c.o.m. 
energies $\sqrt{s}\approx$~14 TeV the rate is lowered by a factor $\approx$~5.

Furthermore, the minimal length acts as a natural regulator\cite{Hossenfelder:2004up} for ultraviolet divergences. 
It therefore
reliably removes an inherent arbitrariness of choice for of cut-off regulators in higher dimensional quantum 
field theories.
 
\section{Discussion}
 
We have worked out the details of an extra dimensional model with a minimal length scale.
The arising modifications have been examined for high precision and for high energy experiments. 
The important conclusion can be drawn that this fundamental limit to the possible resolution of
space time might prevent a further progress in high energy colliders once the new fundamental
scale is reached.
We have shown that, within the model of {\sc LXD}s, not only effects of quantum 
gravity occur at lowered energies but so do the effects of the minimal length scale.  
The minimal length has to be included into the model of {\sc LXD}s for self-consistence. 

\section*{Acknowledgments}

{This work
was supported by a fellowship within the Postdoc-Programme of the German Academic 
Exchange Service 
({\sc DAAD}) and NSF PHY/0301998.}


\end{document}